\documentclass{ws-procs9x6}

\def\Journal#1#2#3#4{{\em #1} {\bf #2}, #4 (#3)}

\begin{document}

\title{Hybrid Mesons from the Lattice}


\author{C. MICHAEL}

\address{Department of Mathematical Sciences, \\
University of Liverpool, \\ 
Liverpool, L69 3BX, UK\\
E-mail: c.michael@liv.ac.uk}

\maketitle

\abstracts{
 I review lattice QCD results for  hybrid mesons,
including  a discussion of their hadronic decays.
 }

\section{Introduction}     
 \label{ex.sect1}

 Quantum Chromodynamics has emerged as the unique theory to 
describe hadronic physics. It is formulated in terms of gluonic and quark
fields. The only free parameters are the scale of the coupling (usually
called $\Lambda_{QCD}$)  and the quark masses defined at some
conventional energy scale.

 Where large momentum transfers occur,  the  effective coupling becomes
weak and a perturbative treatment is valid: in  this domain the theory
has been tested directly by experiment. However, because the effective
coupling is weak for these processes that can be  described by
perturbation theory, they are necessarily not the dominant hadronic 
processes. A typical hadronic process will involve small momentum 
transfers and so has to be treated non-perturbatively.

 In this non-perturbative r\'egime, the description of hadrons is quite
far removed  from the description of the gluonic and quark fields in the
QCD Lagrangian. Because only colour-singlet states survive, the hadrons
are  all composites of quarks and gluons. One example emphasises this:
the  nucleon has a mass which is very much greater than the sum of the 
quark masses of the three valence quarks comprising it. This extra mass 
comes from the gluonic interactions of QCD. Another way to view this is
that  the na\"{\i}ve quark model is a useful phenomenological tool but
has constituent quarks with masses much greater than the QCD masses (ie
masses  as defined in the Lagrangian). It is important to understand why
this  is approximately what QCD requires and to find where QCD departs
from  the na\"{\i}ve quark model.

 One way to characterise the manner in which QCD goes beyond the na\"{\i}ve
 quark model is through the concept of exotic states. Here exotic is
taken to mean  \lq not included in the na\"{\i}ve quark model\rq. 
 In order to discuss exotic states, we need to summarise what the
na\"{\i}ve quark  model contains. Basically the degrees of freedom are the
valence quarks  (ie quark-antiquark for a meson and 3 quarks for a
baryon) with masses and interactions  given by some effective
interaction. The consequences of this are that only certain $J^{PC}$
values  will exist and that the number of states with different quark
flavours is  specified. So, concentrating on mesons made of the three
flavours of light quarks ($u,\ d,\ s$),  one expects a nonet of mesons
with the flavours ($\bar{u} d, \  \bar{d} u,\  \bar{u} u \pm \bar{d} d,\
\bar{s} s,\ \bar{u} s,\ \bar{d} s,\   \bar{s} u,\ \bar{s} d$).  This is
indeed  what is found for vector mesons ($\rho,\ \omega,\ \phi,\ K^{*}$). 
It is also possible within the quark model for the flavour-singlet states 
($\bar{u} u + \bar{d} d,\  \bar{s} s$) to mix, as found for the pseudoscalar 
mesons. What would be exotic is for a tenth state to exist. 
  For mesons with orbital angular momentum $L$  between the quark and
antiquark the  allowed $J^{PC}$ values are shown below.  Thus spin-parity 
combinations such as $0^{--}, 0^{+-}, 1^{-+}, 2^{+-}$ are termed
spin-exotic  since they cannot be made from a quark plus antiquark
alone. 

\begin{table}
\begin{center}
\begin{tabular}{|c|c|ccc|} \hline
  $L$ &   $J^{PC}$ &$J^{PC}$ &$J^{PC}$ &$J^{PC}$ \\ \hline
0 & $0^{-+}$ &          & $1^{--}$ &           \\
1 & $1^{+-}$ & $0^{++}$ & $1^{++}$ & $ 2^{++}$ \\ 
2 & $2^{-+}$ & $1^{--}$ & $2^{--}$ & $ 3^{--}$ \\ 
\hline
\end{tabular}
\end{center}
\end{table}

It has been a considerable challenge to  build a machinery that allows
non-perturbative calculations in QCD with all systematic  errors determined.
 The most controlled  approach to non-perturbative QCD is via
lattice techniques in which space-time is discretized and time is taken
as  Euclidean. The functional integral is then evaluated numerically
using  Monte Carlo techniques. 

  Lattice QCD needs as input the quark masses and an overall scale
(conventionally  given by $\Lambda_{QCD}$). Then any Green function can
be evaluated by taking an average of suitable combinations of the
lattice fields in the vacuum samples. This allows masses to be studied 
easily and matrix elements (particularly those of weak or
electromagnetic currents)  can be extracted straightforwardly.
  Unlike experiment, lattice QCD can vary the quark masses and can also 
explore different boundary conditions and sources. This allows a wide
range of  studies which can be used to diagnose the health of
phenomenological models as well as casting light on experimental data.

Here we will concentrate on lattice results for the spectrum of spin-exotic 
mesons: which are commonly called ``hybrid mesons'' since they must have 
additional degrees of freedom (such as gluonic) to achieve those 
spin-parity values.

One limitation of the  lattice approach  to QCD is  in exploring
hadronic decays because the  lattice, using Euclidean time, has
important contributions from low lying thresholds~\cite{cmdecay} which 
can obstruct the study of decay widths.  The finite spatial size of the
lattice implies that two-body  states are actually discrete. By
measuring their energy very precisely as the spatial  volume is varied,
it is possible~\cite{luscher} to extract the scattering phase shifts and
hence decay properties. For on-shell transitions,  it is  possible to
estimate hadronic transition strengths  more directly~\cite{rhodecay}
and this has recently been checked for the case of  $\rho$ meson decay
to $\pi \pi$.
 This approach has been used to explore hybrid meson  decay rates.

 \section{Hybrid Mesons}
 \label{ex.sect3}

 A hybrid meson is a meson in which the gluonic degrees of freedom are 
excited non-trivially. The most direct sign of this would be  a
spin-exotic meson, since that could not be created from  a $q \bar{q}$
state with unexcited glue. A spin-exotic meson  could, however,  be  a
$q \bar{q}q \bar{q}$ or meson-meson state and that possibility will be
discussed.
 I first discuss hybrid mesons with static heavy quarks where the
description  can be thought of as an excited colour string.   The
situation  concerning light quark hybrid mesons is then summarised 

\subsection{Heavy quark hybrid mesons}
 \label{ex.sect3.1}

\begin{figure}[bt] 
\resizebox{0.8\textwidth}{!}{  
  \includegraphics{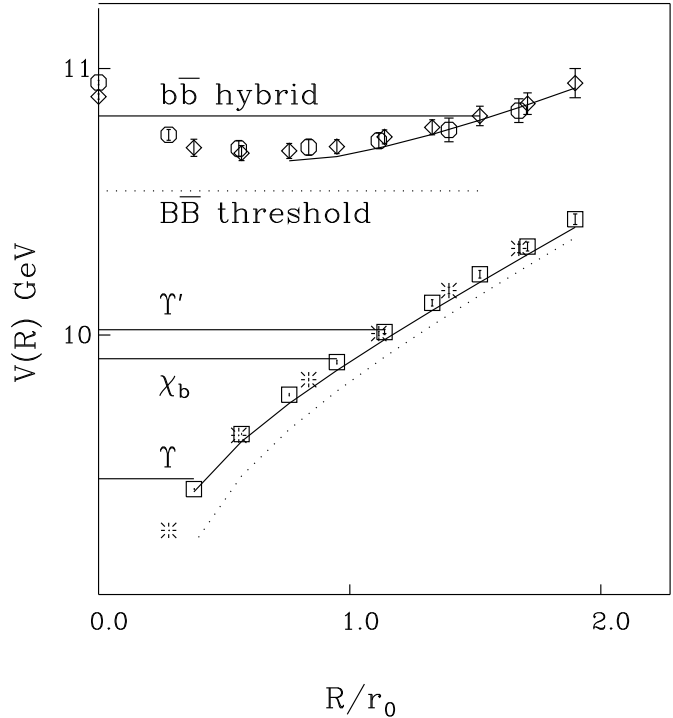}
 }
 \caption{ The potential energy between static quarks at separation $R$
(in units of  $r_0 \approx 0.5$ fm)~{\protect\cite{pm}}. The symmetric
gluonic field  configuration is shown by the lower points while the
$\Pi_u$ excited gluonic configuration is shown above. The energy levels
in these potentials for  $b$ quarks are shown using the adiabatic
approximation. 
   }
\end{figure}

 This topic has a long history: the first paper~\cite{liv} was published
20 years ago. The basic ideas have not changed, and I summarise them
here.

Consider $Q \bar{Q}$ states with static quarks  in which the gluonic
contribution may be excited. We  classify the gluonic fields according
to the symmetries of the system.  This discussion is very similar to the
description of electron wave functions in  diatomic molecules. The
symmetries are  (i) rotation around the separation axis $z$ with
representations labelled by $J_z$ (ii) CP with representations labelled
by $g(+)$ and $u(-)$ and (iii) C${\Re}$. Here  C interchanges $Q$ and
$\bar{Q}$, P is parity and ${\Re}$ is a rotation  of $180^0$ about the
mid-point around the $y$ axis. The C${\Re}$ operation is only relevant
 to classify states with $J_z=0$. The convention is to label states of
$J_z=0,1,2$ by $ \Sigma, \Pi, \Delta$  respectively. The ground state
($\Sigma^+_g$) will have $J_z=0$ and $CP=+$.

 The exploration of the energy levels  of other representations has a
long history in lattice studies~\cite{liv,pm}. The first excited state
is found  to be the $\Pi_u$.  This can be visualised  as the symmetry of
a string bowed out in the $x$ direction minus the same  deflection in
the $-x$ direction (plus another component of  the two-dimensional
representation with the transverse direction $x$ replaced by $y$),
corresponding to flux  states from a lattice  operator which is the
difference of U-shaped paths from quark to antiquark of the form $\,
\sqcap - \sqcup$. 

The picture of the gluon flux between the static  quarks suggests that
the excited states of this string may approximate the excited potentials
found from the lattice. In the simplest string  theory, the first
excited level has $\Pi_u$ symmetry and is at energy $\pi/R$ above the
ground state.  This is indeed  approximately valid and a closer
approximation  is to use a relativistic version~\cite{phm} (namely
$E_m(R)=(\sigma^2 R^2+2\pi\sigma (m-1/12))^{1/2}$ for the $m$-th level),
see also ref.~\cite{jkm} for a recent comparison of this expression.

Recent lattice studies~\cite{jkm}  have used an asymmetric space/time
spacing which enables excited states to be  determined comprehensively.
 These results confirm the finding that 
the $\Pi_u$ excitation is the lowest lying and hence of most relevance 
to spectroscopy.

 From the potential corresponding to these excited gluonic states, one
can  determine the spectrum of hybrid quarkonia using the Schr\"odinger
equation in the Born-Oppenheimer approximation.  This approximation will
be good if the heavy quarks move very little in the  time it takes for
the potential between them to become established. More  quantitatively,
we require that the potential energy of gluonic excitation is much
larger than the typical energy of orbital or radial excitation.  This is
indeed the case~\cite{liv}, especially for $b$ quarks. Another nice
feature of this approach is that the  self energy of the static sources
cancels in the energy difference between this  hybrid state and the
$Q \bar{Q}$ states. Thus the lattice approach gives directly the
excitation energy  of each gluonic excitation.

  The $\Pi_u$ symmetry state corresponds to  excitations of the gluonic
field in quarkonium called magnetic (with $L^{PC}=1^{+-}$) and
pseudo-electric (with $1^{-+}$) in contrast to the usual  P-wave orbital
excitation which has $L^{PC}=1^{--}$. Thus we expect different quantum
number assignments from those of the gluonic ground state. Indeed
combining with the heavy quark spins, we get a degenerate  set of 8
states:

\begin{table}
\begin{center}
\begin{tabular}{|c|c|ccc|} \hline
  $L^{PC}$ &   $J^{PC}$ &$J^{PC}$ &$J^{PC}$ &$J^{PC}$ \\ \hline
$1^{-+}$ & $1^{--}$ & $0^{-+}$ & $1^{-+}$ & $ 2^{-+}$ \\ 
$1^{--}$ & $1^{++}$ & $0^{+-}$ & $1^{+-}$ & $ 2^{+-}$ \\ 
\hline
\end{tabular}
\end{center}
\end{table}

\noindent  Note that of these,  $J^{PC}=  1^{-+},\ 0^{+-}$ and  
$2^{+-}$  are spin-exotic and hence will not mix with $Q\bar{Q}$ states.
They thus form a very attractive goal for experimental searches for
hybrid  mesons.

 The eightfold degeneracy of the static approach will be broken by 
various corrections. As an example, one of the eight degenerate  hybrid
states is a pseudoscalar with the heavy quarks in a spin triplet.  This
has the same overall quantum numbers as the S-wave  $Q \bar{Q}$ state
($\eta_b$) which, however, has the heavy quarks in a spin singlet. So
any  mixing between these states must be mediated by spin dependent
interactions.  These spin dependent interactions will be smaller for
heavier quarks. It is  of interest to establish the strength of these
effects for $b$ and $c$ quarks. Another topic of interest is the
splitting  between the spin exotic hybrids which will come from the
different  energies  of the magnetic and pseudo-electric gluonic
excitations.

 One way to go beyond the static approach is to use the NRQCD
approximation which then enables  the spin dependent effects to be
explored.  One study~\cite{jkm} finds that the  $L^{PC}=1^{+-}$ and
$1^{-+}$ excitations  have no statistically significant splitting 
although the $1^{+-}$  excitation does lie a little lighter. This would
imply, after adding in heavy quark spin, that  the $J^{PC}=1^{-+}$
hybrid was the lightest spin exotic. Also a relatively large spin
splitting was found~\cite{cppacs} among the triplet states considering,
however,   only  magnetic gluonic excitations.
 Another study~\cite{hyb-mix} explores the mixing of non spin-exotic
hybrids  with regular quarkonium states via a spin-flip interaction 
using lattice NRQCD.

 Confirmation of the ordering of the spin exotic states also comes from
 lattice studies with propagating quarks~\cite{livhyb,milc,sesamhyb}
which  are able to measure masses for all 8 states. I discuss that
evidence in more detail below.

 Because of the similarity of the lightest hybrid wavefunction with 
that of the 2S state (which has $L=1$), it is convenient to 
quote mass differences between these states.  
Within the quenched approximation,  the lattice evidence  for
$b\bar{b}$ quarks points to a  lightest hybrid spin exotic with
$J^{PC}=1^{-+}$ at an energy given by $(m_H-m_{2S})r_0$ =1.8 (static
potential~\cite{pm}); 1.9 (static potential~\cite{jkm},
NRQCD~\cite{cppacs}); 2.0 (NRQCD~\cite{jkm}). These results can be
summarised as       $(m_H-m_{2S})r_0=1.9 \pm 0.1$. 
 Using the experimental mass of the $\Upsilon(2S)$, this implies that
the lightest spin exotic  hybrid is at $m_H=10.73(7)$ GeV including a
10\% scale error.  Above this energy there will be many more hybrid 
states, many of which will be spin exotic.

 The results from a study  with $N_f=2$ flavours of sea-quarks show very
little change in the static potential as shown by SESAM~\cite{sesam} and
as illustrated in fig.~\ref{ex.hdecay} and also relatively little change
in NRQCD determinations~\cite{cppacs} of mass ratios such  as
$(m_H-m_{2S})/(m_{1P}-m_{1S})$. Expressed in terms of $r_0$  (using
$r_0=1.18/\sqrt{\sigma}$) this gives $(m_H-m_{2S})r_0=2.4(2)$, however. 
This is significantly larger than the quenched result and, using the
$1P-1S$  mass difference to set the scale, yields a
prediction~\cite{cppacs} for the lightest hybrid mass  of 11.02(18) GeV.

\subsection{Hybrid meson decays}
 \label{ex.sect3.2}

 Within this static quark framework, one can explore the decay
mechanisms.  One special feature is that the symmetries of the quark and
colour fields about the static quarks must be preserved exactly in
decay, hence the light quark-antiquark pair produced must respect these
symmetries.   This has the consequence that the decay from a $\Pi_u$
hybrid state to the open-$b$ mesons ($B \bar{B},\   B^* \bar{B},\  B
\bar{B^*},\ B^* \bar{B^*}$) will be forbidden~\cite{hf8,hdecay} if the 
light quarks in the $B$ and $B^*$ mesons are in an S-wave relative to
the heavy quark (since the final state will have the light quarks in
either a triplet  with the wrong $CP$ or a singlet with the wrong $J_z$
where $z$ is the interquark axis for the heavy quarks). The decay to
$B^{**}$-mesons with light quarks in a P-wave is allowed by symmetry but
not energetically.

\begin{figure}[th]
\resizebox{0.8\textwidth}{!}{  
  \includegraphics{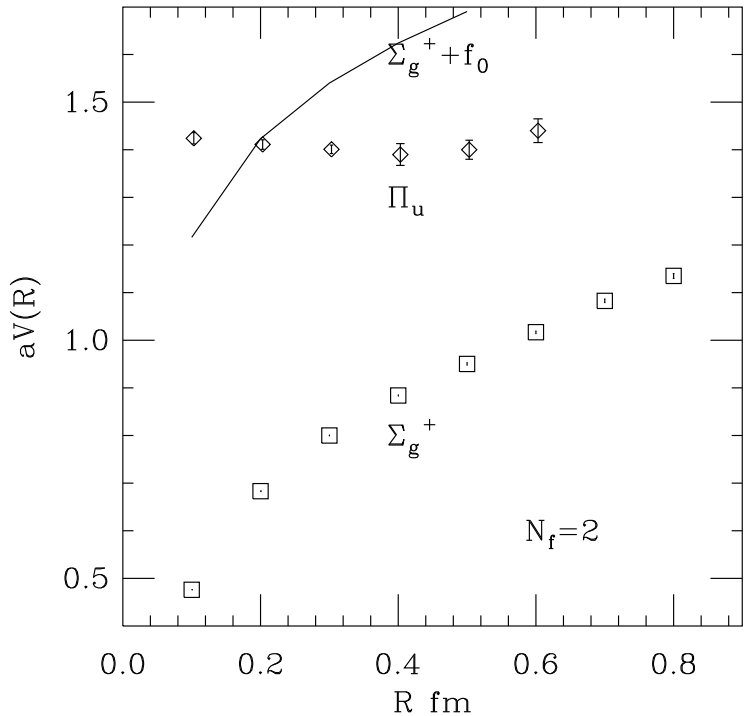} }
 \caption{The potential energy $V(R)$ (in lattice units with a=0.0972 fm)
versus quark separation $R$ in fm  for  2 flavours of sea quark. The 
energies are given~{\protect\cite{hdecay}}  for the ground state and
first excited gluonic state  and for the two body state of ground state
potential plus scalar meson  ($f_0$) in a P-wave with the minimum
non-zero momentum. The on-shell transition can be evaluated when $R
\approx 0.2$ fm.
 }
 \label{ex.hdecay}
\end{figure}

In the heavy quark limit, the only allowed decays are when the hybrid
state de-excites to a  non-hybrid state with the emission of a light
quark-antiquark pair. Since the  $\Pi_u$ hybrid state has the heavy
quark-antiquark in a triplet P-wave state,  the resulting non-hybrid
state must also  be in a triplet P-wave since the heavy  quarks do not
change their state in the limit of very heavy quarks. Thus the decay for
$b$ quarks will be  to $\chi_b +M$ where $M$ is  a light quark-antiquark
meson in a flavour singlet. This proceeds by a disconnected  light quark
diagram and it would be expected~\cite{vall} that the scalar or
pseudoscalar meson  channels are the most important (ie they have the
largest relative OZI-rule  violating contributions).  
 This transition can be estimated on a lattice when the intial and final
energies  are similar. This is the case~\cite{hdecay} for the $\Pi_u$
de-excitation to ground state  gluonic field plus $f_0$ meson when the
interquark separation is around  0.2 fm which allows a lattice
evaluation of the  hadronic transition strength - see
fig.~\ref{ex.hdecay}.  Indeed the dominant mode (with a width of around
100 MeV) is found to  be with $M$ as a scalar meson, namely $H \to
\chi_b + f_0$, whereas when $M$ is an $\eta$ or $\eta'$ meson the
transition strength is less that a few MeV.
 There will be modifications to this analysis coming from corrections to
the heavy quark limit  (of order $1/m_Q$ where $m_Q$ is the heavy quark
mass) which might allow hybrid meson transitions to  $B \bar{B}$, etc,
but these have not been evaluated yet.

 In this heavy quark (or static) limit,  the spin-exotic and non
spin-exotic hybrid  mesons are degenerate. For the latter, however,  the
interpretation of any observed states is less clear cut, since they 
could be conventional quark antiquark states. Moreover, the non
spin-exotic  hybrid mesons can mix directly (ie without emission of any
meson $M$) with conventional quark antiquark states once  one takes into
account corrections (of order $1/M_Q$) to the static approximation 
applicable for  heavy quarks with physical masses. 

 It is encouraging that the decay width comes out as relatively 
small, so that the  spin-exotic hybrid states should show up 
experimentally as sufficiently narrow resonances to be detectable.
 This decay analysis does not take into account heavy quark motion or
spin-flip  and these effects will be significantly more important for
charm quarks than for $b$-quarks.

\subsection{Light quark hybrid mesons}
 \label{ex.sect3.3}

 I now  focus on lattice results for hybrid mesons made from light
quarks using fully relativistic propagating quarks.  There will be no
mixing with $q \bar{q}$ mesons for  spin-exotic hybrid mesons  and these
are of special interest. The first study of this area was by the  UKQCD
Collaboration~\cite{livhyb} who used operators motivated by the  heavy
quark studies referred to above to study all 8 $J^{PC}$ values coming
from $L^{PC}=1^{+-}$ and $1^{-+}$ excitations. The  resulting mass
spectrum  gives the $J^{PC}=1^{-+}$ state as the lightest spin-exotic
state. Taking account of the systematic scale errors in the lattice
determination, a  mass of 2000(200) MeV is quoted for this hybrid meson
with $s \bar{s}$ light quarks. Although not directly measured, the
corresponding light quark hybrid meson would be expected to be around
120 MeV lighter.

A second lattice group has also evaluated hybrid meson spectra with
propagating quarks  from quenched lattices. They obtain~\cite{milc}
masses of the $1^{-+}$ state with statistical and various systematic
errors of  1970(90)(300) MeV, 2170(80)(100)(100) MeV and 4390(80)(200)
MeV for $n \bar{n}$,  $s \bar{s}$ and $c \bar{c}$ quarks respectively.
For the  $0^{+-}$ spin-exotic state they have a noisier signal but
evidence that it is heavier. They also explore mixing matrix elements
between spin-exotic hybrid  states and 4 quark operators.

 The first analysis~\cite{sesamhyb} to determine the hybrid meson
spectrum using  full QCD used Wilson quarks. The sea quarks used had
several different masses and an extrapolation  was made to the limit of
physical sea quark masses, yielding a mass of 1.9(2) GeV for the
lightest  spin-exotic hybrid meson, which again was found to be the
$1^{-+}$. In principle this  calculation should take account of sea
quark effects such as the mixing  between such a hybrid meson and $q
\bar{q} q \bar{q}$ states such as $\eta \pi$, although it is possible
that the sea quark  masses used are not light enough to explore these
features.

 A  recent dynamical quark study from 2+1 flavours of improved staggered
quarks has also produced results~\cite{milc2}. They also compare their
results with quenched calculations and find  no significant difference,
except that the ambiguity in fixing the lattice  energy scale is better
controlled in the dynamical simulation since different reference
observables are closer to experiment. Their summary result for the
$1^{-+}$ hybrid with strange quarks is  $2100 \pm 120$ MeV, in agreement
with earlier results. They note that the  energies of  two-meson states
(such as $\pi + b_1$ or $K + K(1^{+})$ ) with the hybrid  meson quantum
numbers are close to the energies they obtain. This suggests that  these
two-particle states, which are allowed to mix in a dynamical quark
treatment, may be  influencing the masses determined. A study of hybrid
meson transitions to two particle states is needed  to illuminate this
area, using techniques such as those used for heavy quark hybrid
decay~\cite{hdecay}  and  decays of light quark  vector 
mesons~\cite{rhodecay}.

The lattice calculations~\cite{livhyb,milc,sesamhyb,milc2,ml} of the
light hybrid spectrum are  in good agreement with each other. They imply
that the natural energy  range for spin-exotic hybrid mesons is around
1.9 GeV. The $J^{PC}=1^{-+}$  state is found to be lightest. It is not
easy to reconcile these lattice results  with experimental
indications~\cite{expt} for resonances at 1.4 GeV and 1.6 GeV,
especially the  lower mass value.  Mixing  with  $q \bar{q} q \bar{q}$
states such as $\eta \pi$ is not included for realistic quark masses in
the  lattice calculations. Such effects of pion loops (both real and
virtual) have been estimated  in chiral perturbation theory based
models~\cite{aa} and they could potentially reconcile some of the
discrepancy between lattice mass estimates (with light quarks which are
too heavy)  and those from experiment. This can be interpreted,
dependent on one's viewpoint,  as either that the lattice calculations 
are incomplete or as an indication that the experimental states may have
an  important meson-meson component in them.

 The light quark technique of using relativistic propagating quarks 
can also be extended to charm quarks, as was note above~\cite{milc}.
 Another group has explored the charm quark hybrid states also using a
fully  relativistic action, albeit with an anisotropic  lattice
formulation~\cite{manke}. Their quenched study is in agreement with the
isotropic lattice result quoted above, finding  a mass value of 
4.428(41) GeV in the continuum limit for the  $1^{-+}$ hybrid where the
scale is set by the $^1P_1 - 1S$ mass splitting (458.2 MeV
experimentally) in charmonium.  Their result is also consistent with
that from NRQCD methods~\cite{cppacs} applied to this case. These 
results all have the usual caveat that in quenched evaluations the
overall mass scale of the energy difference  from the $1S$ state at
3.067 GeV is uncertain to  10\% or so (for example the
$(2S-1S)/(^1P_1-1S)$ is  found to be 15\% higher than experiment) which
is a major source of systematic error (approximately $\pm 140$ MeV). 
They  also produce estimates for other charmonium spin-exotic states: 
$0^{+-}$ at 4.70(17) GeV and $2^{+-}$ at 4.89(9) GeV. The $0^{--}$ state
is not resolved.

Thus  masses near 4.4GeV  are found for the  charmonium $1^{-+}$ state
using relativistic quarks. The non-relativistic approach using  NRQCD is
expected to have big systematic errors for quarks as light as charm, but
results~\cite{cppacs} do agree with this value.
 The heavy quark effective theory approach has a leading term which
corresponds to a static heavy quark,  resulting in an estimate~\cite{pm}
of  the spin-exotic charm  state mass of 4.0 GeV.  Here again the 
systematic error is potentially large for charm quarks.

\section{Conclusions and Outlook}
 \label{ex.sect5}

 For hybrid mesons, there will be no mixing with $q \bar{q}$ for 
spin-exotic states and these are the most useful predictions. The
$J^{PC}=1^{-+}$ state is expected in the range 10.7 to 11.0 GeV for $b$ quarks,
 2.0(2) GeV for $s$ quarks and 1.9(2) GeV 
for $u,\ d$ quarks. Mixing of spin-exotic hybrids with
$q\bar{q}q\bar{q}$ or equivalently with meson-meson  is  allowed and
will modify the  predictions from the quenched approximation.
 A first lattice study has been made of hybrid meson decays. For heavy 
quarks, the dominant mode is string de-excitation to $\chi + f_0$ where
$f_0$ is  a flavour singlet scalar meson (or possibly two pions in this
state). The magnitude of the decay rate is found to be of order 100 MeV,
 so this decay mode should  still leave a detectably narrow resonance to
 be observed. It will be possible to explore hadronic decays for 
light-quark hybrid mesons in the near future, and this may help to 
elucidate the nature of the experimental spin-exotic signals found  at
1.4 and 1.6 GeV.

%
%
%
%

\end{document}